\documentclass[conference]{IEEEtran}
\usepackage[left=0.635in,right=0.635in,top=0.75in,bottom=1in]{geometry}
\makeatletter
\def\ps@headings{%
	\def\@oddhead{\mbox{}\scriptsize\rightmark \hfil \thepage}%
	\def\@evenhead{\scriptsize\thepage \hfil \leftmark\mbox{}}%
	\def\@oddfoot{}%
	\def\@evenfoot{}}
\makeatother
\usepackage{listings}
\usepackage{algorithm, algpseudocode}
\usepackage{url}
\usepackage{pgfplots}
\usepackage{tabularx}
\usepackage{amsthm}
\usepackage{comment}
\theoremstyle{definition}
\newtheorem{definition}{Definition}[section]
\makeatletter
\algrenewcommand\ALG@beginalgorithmic{\footnotesize}
\makeatother
\usepackage{listings}
\definecolor{codegreen}{rgb}{0,0.6,0}
\definecolor{codegray}{rgb}{0.5,0.5,0.5}
\definecolor{codepurple}{rgb}{0.50,0,0.80}
\definecolor{backcolour}{rgb}{0.95,0.95,0.92}
\definecolor{lbcolor}{rgb}{0.9,0.9,0.9}

\lstdefinestyle{mystyle}{
	backgroundcolor=\color{lbcolor},   
	commentstyle=\color{codegreen},
	keywordstyle=\color{magenta},
	numberstyle=\tiny\color{codegray},
	stringstyle=\ttfamily,
	basicstyle=\footnotesize,
	breakatwhitespace=false,         
	breaklines=true,                 
	captionpos=b,                    
	keepspaces=true,                 
	numbersep=5pt,                  
	showspaces=false,                
	showstringspaces=false,
	showtabs=false,                  
	tabsize=2
}

\ifCLASSOPTIONcompsoc
\usepackage[nocompress]{cite}
\else
\usepackage{cite}
\fi
\ifCLASSINFOpdf
\else
\fi
\begin{document}
	%
	\title{SDFW: SDN-based Stateful Distributed Firewall}
	%
	%
	%
	%
	
	\author{
		\IEEEauthorblockN{Ankur Chowdhary, Dijiang Huang, Adel Alshamrani and Abdulhakim Sabur}
		\IEEEauthorblockA{Arizona State University
			\\\{achaud16, dijiang, aalsham4, asabur\}@asu.edu}
		
		\IEEEauthorblockN{Myong Kang, Anya Kim and Alexander Velazquez}
		\IEEEauthorblockA{Naval Research Lab
			\\\{myong.kang, anya.kim, alexander.velazquez\}@nrl.navy.mil}		
	}
	%
	%

	\markboth{Journal of \LaTeX\ Class Files,~Vol.~14, No.~8, August~2015}%
	{Shell \MakeLowercase{\textit{et al.}}: Bare Advanced Demo of IEEEtran.cls for IEEE Computer Society Journals}
	%



	\IEEEtitleabstractindextext{%
		\begin{abstract}
			SDN provides a programmable command and control networking system in a multi-tenant cloud network using control and data plane separation. However, separating the control and data planes make it difficult for incorporating some security services (e.g., firewalls) into SDN framework. Most of the existing solutions use SDN switches as packet filters and rely on SDN controllers to implement firewall policy management functions, which is impractical for implementing stateful firewalls since SDN switches only send session's initial packets and statistical data of flows to their controllers. For a data center networking environment, applying a Distributed  FireWall (DFW) system to prevent attacker's lateral movements is highly desired, in which designing and implementing an SDN-based Stateful DFW (SDFW) demand a scalable distributed states management solution at the data plane to track packets and flow states. Our performance results show that SDFW achieves scalable security against data plane attacks with a marginal performance hit $\sim$ 1.6\% reduction in network bandwidth. 
			
		\end{abstract}
		
		}

	\maketitle

	\IEEEdisplaynontitleabstractindextext

	%
	\IEEEpeerreviewmaketitle

	\ifCLASSOPTIONcompsoc
	\IEEEraisesectionheading{\section{Introduction}\label{sec:introduction}}
	\else
	\section{Introduction}
	\label{sec:introduction}
	\fi
	\noindent Software Defined Networking (SDN)~\cite{kreutz2015software} simplifies networking management by decoupling control plane and data plane. The SDN controller can dynamically configure multiple physical or virtual network switches. The lack of built-in security in the SDN limits its adoption, as reported some campus adopters~\cite{junipersdn}. Although the centralized design is an import characteristic of SDN framework, it can introduce security challenges such as denial-of-service (DoS) attacks, targeted at SDN controller and OpenFlow switches~\cite{scott2013sdn}. An important role of DFW is the prevention of lateral movement of an attacker as described in microsegmentation framework~\cite{mammela2016towards}, which is not implemented by current SDN firewalls. In particular, this research focuses on the following two research issues 
	
	\noindent \textbf{SDN-based  Firewall Issues: }There are two main issues we identified in current SDN-based firewall architecture. 1) Most of the existing firewall architectures such as Flowguard~\cite{hu2014flowguard}, FortNOX~\cite{porras2012security},~\cite{suh2014building},~\cite{zerkane2016software} are centralized in nature. In a large cloud network, SDN controller performance can become extremely slow in the centralized architecture. 2) There is no support for packet state maintenance and multi-tenancy in most of these works, as noted by Dixit \textit{et al}~\cite{dixit2018challenges}. In the absence of state information, it is difficult to discover attacks originating in SDN-data plane~\cite{gao2018security}. Thus, a stateful firewall is necessary to provide a granular security in SDN-environment. The challenge, however, with a centralized stateful firewall, is that controller receives a large volume of state-based traffic, which can become a performance bottleneck. So we need a distributed stateful-firewall (SDFW) to provision granular security in a scalable fashion. Using an SDFW architecture, the distributed firewall can be easily managed in a cloud network.
	
	With these security challenges and design goals, discussed above in mind, and realizing the need for an automated-security management framework, we designed \textit{SDFW}. The key contribution of this research work are as follows:
	\begin{itemize}
		\item The SDFW firewall in our framework is utilized to construct OpenFlow rules, which are implemented on the switches using a stateful distributed firewall (SDFW) framework. Using the SDFW scales well on a large network with limited performance impact  $\sim$ 1.6\% reduction in network bandwidth.
		\item SDFW identifies the lateral movement of the attacker and implements SDN based security countermeasures to prevent the attack propagation in a multi-tenant cloud network.

	\end{itemize}
	
	\section{Background and Motivation}
	\subsection{SDN and OpenFlow}
	\begin{figure}[ht!]
		\centering
		\includegraphics[width=0.50\textwidth]{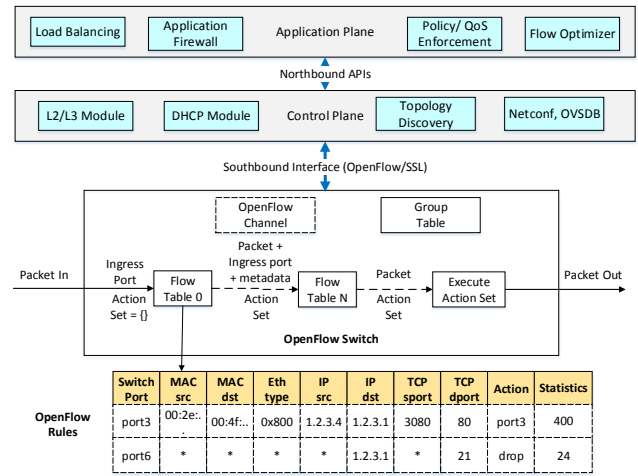}
		\caption{OpenFlow Rule Format}
		\label{fig:1}  
	\end{figure}
	\noindent OpenFlow is one of the most popular protocols, which allows the incorporation of SDN capabilities in the cloud network. OpenFlow switches consist of flow tables, which run at line rate to allow traffic between data plane devices. Figure~\ref{fig:1} illustrates different planes of SDN, and how SDN controller can insert flow rules into flow tables.
	
	\begin{definition}\label{def1}
		\textbf{OpenFlow Rule:} A flow table F of an OpenFlow switch, can have rules, $\{r_1, r_2,.., r_n\}$ Each rule consists of layer 2-4 packet header fields, protocol (TCP/UDP/FTP), action-set associated with the rule, rule priority, and statistics. We define the flow rule using tuple $r_i$ = ($p_i$, $\rho_i$, $h_i$, $a_i$, $s_i$), where a) $p_i$ denotes rule priority, b) $\rho_i$ denotes the protocol of the incoming traffic (TCP/UDP) c) $h_i$ depicts the packet header, d) $a_i$ is the action associated with the rule, e) $s_i$ represents the statistics associated with the rule. 
		
		The flow rule header space $h_i$, consists of physical port of incoming traffic $\delta_i$, source and destination hardware address, i.e., ${\alpha_s}_i, {\alpha_d}_i$, source and destination IP address, ${\beta_s}_i, {\beta_d}_i$, source and destination port address, ${\gamma_s}_i, {\gamma_d}_i$. Packet header can be defined by the tuple $h_i$ = ($\delta_i$, ${\alpha_s}_i, {\alpha_d}_i$, ${\beta_s}_i, {\beta_d}_i$, ${\gamma_s}_i, {\gamma_d}_i$). Rule statistics $s_i$, comprises of both flow duration and number of packets/bytes for each flow rule $s_i = (d_i, b_i)$.
	\end{definition}
	
	\noindent The flow rules, generally presented in the Figure~\ref{fig:1} can be used to block traffic from a network segment, using SDN based centralized firewall architecture~\cite{dixit2018challenges}. However, the attacks originating in data-plane which rely on connection information, go undetected using the centralized firewall architecture.  
	
	\subsection{Need for Stateful-Distributed Firewall}
	\textbf{Firewall:} is a collection of components, interposed between two networks, that filters the traffic between them according to some security policy. If we consider the modern data-centers as a use-case, the scope of security enforcement offered by a traditional firewall is limited to north-south traffic, i.e., firewall serves as a sentry between trusted and untrusted networks. Once the attacker has managed to breach the security restrictions at the network edge, he can laterally move inside the network (east-west traffic), exploiting key resources, virtually unchecked.  The volume of east-west traffic in the data center environment is around 76\%, as compared to north-south traffic - 17\%~\cite{east-west}.  
	
	\textbf{Stateful Firewall: }is responsible for packet filtering by tracking the state of network connections. The TCP connections have three major states, connection establishment, usage, and termination. The firewall normally utilizes a state-table to track the bidirectional connection between hosts and blocks the packets that deviate from expected state~\cite{wack2002guidelines}. The application firewall that performs fine-grained analysis such as stateful protocol analysis and deep-packet-inspection (DPI), has not been considered in SDFW framework in its current version. 
	
	\begin{figure}[ht!]
		\centering
		\includegraphics[width=0.50\textwidth]{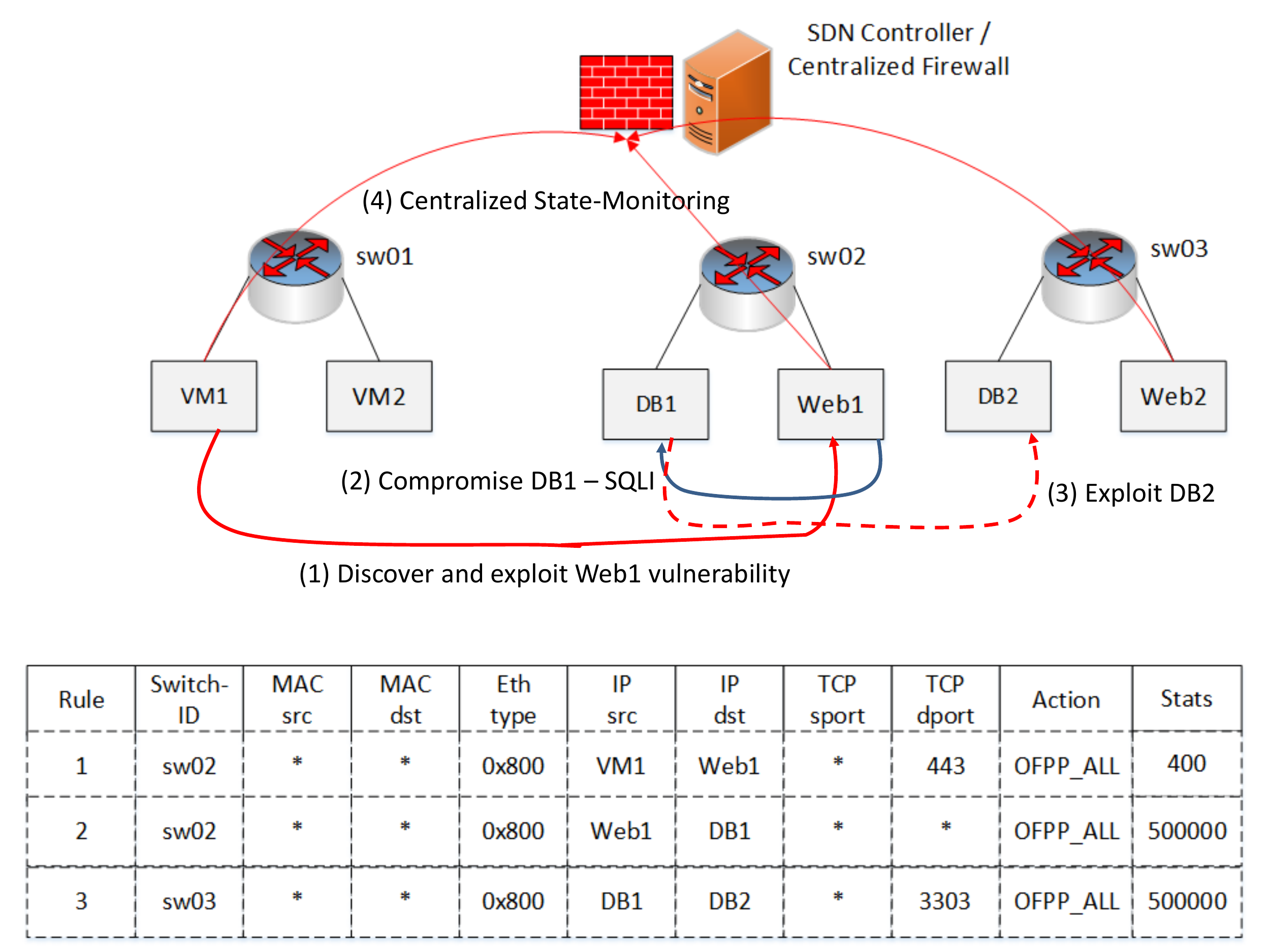}
		\caption{Security Issues in a centralized firewall}
		\label{fig:3}
		\vspace{-1.5em}
	\end{figure}
	
	\noindent \textbf{Scenario 1: Stateless-centralized Firewall} Figure~\ref{fig:3}, shows three OpenFlow switches connected to centralized SDN controller, with Firewall functionality. The traffic between certain VMs across the networks has been allowed, using OpenFlow rules shown in the flow table. Suppose if there is a security vulnerability on Web1, DB1, and, DB2, and the attacker is located on VM1. Although the network traffic is allowed between VM1-Web1, the stateless firewall cannot inspect the state of the network connection. Using a stateless firewall alone, the connection information used by an attacker to mount a multi-hop attack will remain undetected.  
	
	\noindent \textbf{Scenario 2: Stateful centralized Firewall} If the SDN controller decides to inspect every single packet, all the network connection traffic will be redirected to SDN controller, as shown in Figure~\ref{fig:3}. The SDN controller may be quickly overwhelmed. Additionally, the attacker can launch control plane saturation attacks if the connection tracking is enforced on the SDN controller, as highlighted by AVANTGUARD~\cite{shin2013avant}. 
	\section{System Architecture and Data Flow}
	\begin{figure}[ht!]
		\centering
		\includegraphics[width=0.50\textwidth]{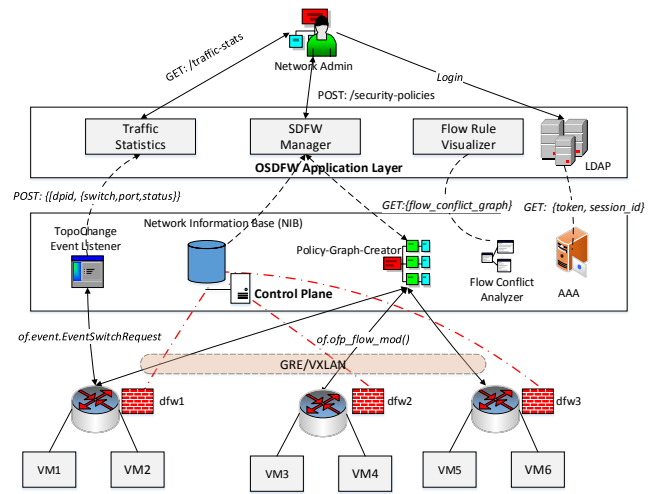}
		\caption{SDFW Architecture}
		\label{fig:2}
	\end{figure}
	
	\noindent The SDFW architecture in Figure~\ref{fig:2} is primarily divided into three planes, i.e., \textit{application plane}, responsible for user-interface, through which the user can enter higher-level security policies, and visualize the state of a distributed firewall. The \textit{control plane} consists of modules, responsible for the translation of higher level security policies, into OpenFlow rules, and identifying any conflicts between OpenFlow rules, and security policies. The \textit{data plane} consists of OpenFlow switches, with state-tracking capability, each OpenFlow switch acts as a firewall module for inspecting traffic between the hosts connected to the switch, and the traffic between switch and control plane. 
	
	\noindent \textbf{SDFW Manager:} checks the status of individual virtual firewalls connected through \textit{Network Information Base} as shown below, and accepts the security policies through the UI written in the PHP-lavarel framework.
	
	\noindent \textbf{Network Information Base (NIB):} acts as a middleware between the application plane implementing distributed firewall policy and local event listeners on each switch. NIB notifies the local-agents on each switch about any new application security policies and maintains synchronization between different agents. NIB has been implemented using Zookeeper~\cite{hunt2010zookeeper}. 
	
	\noindent \textbf{Policy-Graph-Creator: }checks the dependencies between requirements of different security policies, and creates end-to-end conflict-free Policy-Graph to direct traffic between different hosts in a data-center.  The control plane utilizes this Policy-Graph to modify the flow rules of OpenFlow tables, using OpenFlow message \textit{$ofp\_flow\_mod()$} and creates an end-to-end traffic flow. This module checks the dependencies between requirements of different policies. The end result of this process is Policy-Graph.
	
	\noindent \textbf{Traffic Statistics:} The controller consists of \textit{TopoChangeEventListener}, which listens on the events such as port status (UP/DOWN), switch status, port information of hosts connected to switches. If there is any topology change, the event listener utilizes a PUSH notification to notify the application plane, which in-turn updates the visualization and traffic statistics.

	\section{Design of Stateful Distributed Firewall}
	
	\noindent The most popular software switch used by OpenFlow protocol Open vSwitch has a capability to track the connection-state of the packet, as well as the features to define the virtual routing domains in the Linux kernel. Some important fields, of the conntrack module, which we will use in the illustrative example have been defined in the Table~\ref{tab:1} below.
	
	\begin{table}[htb]
		\centering
		\caption{OVS Conntrack Fields}
		\label{tab:1}
		\begin{tabular}{ p{2cm}|p{5.5 cm} }
			\hline  
			\textbf{Field} &   \textbf{Description}\\
			\hline 
			\hline 
			$ct\_state$ & State of the connection tracking module, +/- is used for 
			specifying set, unset. Examples - +new, +esttablished, +trk. \\  
			\hline            
			$ct\_zone$ & independent connection tracking context, set by action. \\
			\hline
			$ct\_nw\_src$, $ct\_nw\_dst$ &  Source and destination IP of connection. \\
			\hline
			$ct\_tp\_src$, $ct\_tp\_dst$ &  Source and destination port of connection. \\
			\hline
			commit & Commit the connection to the connection tracking module. \\
			\hline
		\end{tabular}                
	\end{table}
	
	\noindent In our distributed firewall design, we leverage the information stored by the OpenFlow table and connection tracking table to identify security violations. We use a \textit{Local DFW Event-Listener}, which keeps track of all the stateful connection events, that happen on the \textit{conntrack} module. The event listener can inspect if the activities are malicious or benign, and take corresponding countermeasures to mitigate the security threats.
	
	\begin{figure}[ht!]
		\centering
		\includegraphics[width=0.50\textwidth]{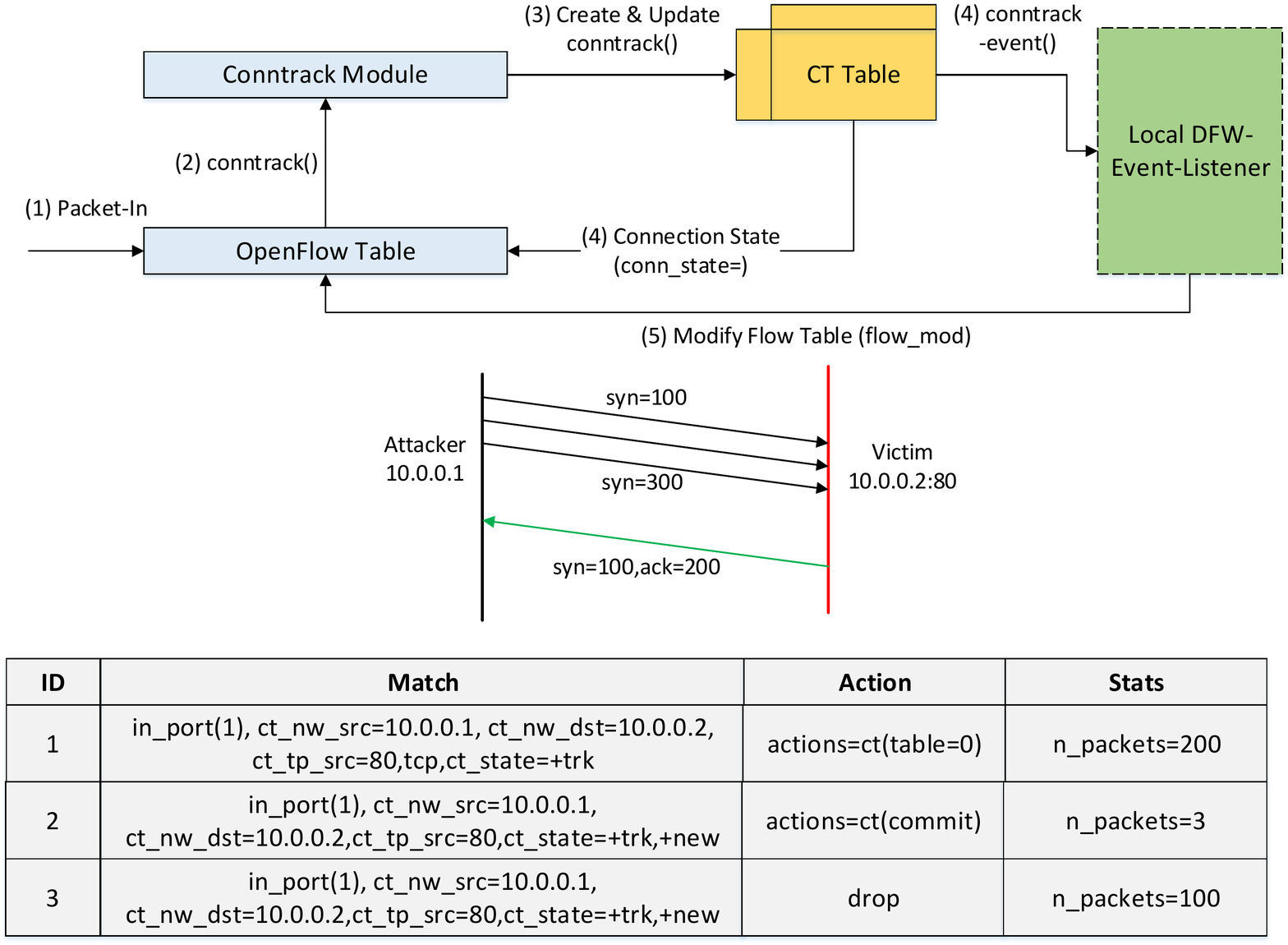}
		\caption{Local State Tracking and Security Analysis, TCP SYN-Flood Attack mitigation.}
		\label{fig:4}
	\end{figure}
	\noindent The Figure~\ref{fig:4}, shows the stepwise handling of security incidents such as \textit{TCP SYN-Flood} attack. Consider the attacker located on the source IP address \textit{10.0.0.1}, sends traffic to the victim on the destination IP address \textit{10.0.0.2}, port 80 - Step (1). 
	
	\begin{lstlisting}
	for i in rang(0,199):
	sendp(Ether()/IP(src="10.0.0.1", dst="10.0.0.2")/TCP(sport=1024, 
	dport=80, flags=0x02, seq=100+i), iface="1")
	\end{lstlisting}
	
	\noindent The \textit{OpenFlow Table} sends the packet to \textit{Conntrack Module} - Step (2). The conntrack module, is responsible for creating and updating connection tracking table - \textit{CT Table} - Step (1). The rule with ID '1', is used by CT table to assign a state=+trk to the new connection, corresponding to the SYN packet (syn=100). Additionally, the OpenFlow module, sends back the SYN-ACK - rule ID '2', to notify the attacker, about the intent for establishing the TCP connection, using response (syn=100,ack=200). 
	
	\noindent The attacker, however, instead of sending ACK=201 corresponding to the (syn=100,ack=200), sends a huge volume of networking traffic as shown using \textit{sendp()} command above. These half-open connections, saturate the network bandwidth of victim '10.0.0.2'.
	
	\noindent \textbf{Local DFW-Event-Listener} also receives the event-notification about the connection state from the CT Table. The module, checks the difference in the \textit{Stats} column for the rules with ID '1', and, '2' (syn=100 to syn=300), in the figure above, and the absence of ACK from the attacker, which is necessary to establish the TCP full connection, and inserts a rule in OpenFlow Table - Step 5, to drop the communication from attacker corresponding to the half-open connections - rule ID '3'. 
	
	
	\subsection{SDN Managed Container Environment: Case Study} 
	\noindent The Linux Containers~\cite{rosen2014linux} have gained popularity in recent times, since they allow quick provisioning of applications, and are easy to manage using the $lxc$ daemon. We created an environment, with about 100 Linux containers of type Ubuntu:16.04 and CentOS 6, which downloads the lxc images for OS, creates the containers and assigns the IP address to containers.
	\begin{lstlisting}
	import lxc
	import sys
	import os
	
	def createContainers():
	for i in range(1,100):
	os.system('lxc-create -t download -n u'+str(i)+ --dist 
	ubuntu --release bionic --arch amd64')
	os.system('sudo lxc-start --name u'+str(i)+' --daemon')
	
	def setVMIP():
	for i in range(1,100):
	os.system('lxc-attach -n u'+str(i)+'-- bash -c \'sudo 
	ip addr add 10.0.3.1'+str(i)+'/24 dev eth0\'')
	os.system('lxc-attach -n u'+str(i)+'-- bash -c \'sudo
	ip route add default via 10.0.3.1\'')
	
	def printContainers():
	for container in lxc.list_containers(as_object=True):
	print(container.name, container.state, container.get_ips(), 
	container.get_cgroup_item("memory.max_usage_in_bytes"))
	
	createContainers()
	setVMIP()
	printContainers()
	\end{lstlisting}

	\noindent We modified the configuration files of each container to attach the containers' port to the Linux bridge 'br100', which we used for the analysis of the target environment as shown below. The 'ovsup' and 'ovsdown' scripts were utilized for attaching and detaching the containers to ovs-bridge when the container is started or stopped. The default configuration option linking the containers to Linux bridge is commented out.

	\begin{lstlisting}
	==============================
	$cat \var\lib\lxc\u1\config
	
	# Distribution configuration
	lxc.include = /usr/share/lxc/config/ubuntu.common.conf
	lxc.arch = linux64
	
	# Container specific configuration
	lxc.rootfs = /var/lib/lxc/u1/rootfs
	lxc.rootfs.backend = dir
	lxc.utsname = ux(1-100)
		
	# Network configuration
	lxc.network.type = veth
	#lxc.network.link = lxcbr0
	lxc.network.script.up = /etc/lxc/ovsup
	lxc.network.script.down = /etc/lxc/ovsdown
	
	
	============ovsup=============
	#!/bin/bash
	BRIDGE="br100"
	ovs-vsctl --may-exist add-br $BRIDGE
	ifconfig $5 0.0.0.0 up
	ovs-vsctl --if-exists del-port $BRIDGE $5
	ovs-vsctl --may-exist add-port $BRIDGE $5
	
	===========ovsdown===========	
	#!/bin/bash	
	BRIDGE="br100"
	ifdown $5
	ovs-vsctl del-port br100 $5                            
	===============================
	\end{lstlisting}
	
	\noindent Each OpenFlow switch runs a \textit{Local DFW-Event-Listener}, a python module, which keeps track of events related to stateful connections. For instance, we used TCP SYN-Flood.py module, which simulates SYN-Flood attack on the container with  u2 - IP (10.0.3.102) to try and send a huge volume of traffic to container u3 - IP (10.0.3.103). With the connection tracking in place, we check the OpenFlow rules, present on the OVS-bridge br100 connecting both hosts. 
	
	\begin{lstlisting}
	
	(1) cookie=0x0, duration=355.095s, table=0, n_packets=1400, n_bytes=75600, 
	nw_src=10.0.3.102,nw_dst=10.0.3.103, reset_counts priority=50,ct_state=-trk,
	tcp,in_port=vethNNE99K actions=ct(table=0)
	
	(2) cookie=0x0, duration=492.169s, table=0, n_packets=1400, n_bytes=75600, 
	nw_src=10.0.3.102,nw_dst=10.0.3.103, reset_counts priority=50,ct_state=+new,
	tcp,in_port=vethNNE99K actions=ct(commit),output:vethMFMXS7
	
	(3) cookie=0x0, duration=72.178s, table=0, n_packets=0, n_bytes=0, 
	nw_src=10.0.3.102,nw_dst=10.0.3.103, reset_counts priority=50,
	ct_state=+est,tcp,in_port=vethNNE99K actions=output:vethMFMXS7
	\end{lstlisting}
	
	\noindent Based on the observation of flow rules, we can see that the attacker, only sends only SYN-packets, $ct\_state = +new$ - rule (2) in the output above, the field $n\_packets=1400$ indicates a huge volume of TCP traffic directed towards the victim. The host 10.0.3.103, sends SYN-ACK, but the attacker, doesn't send back 'ACK', which can lead to state-transition, i.e., $ct\_state=+est$, thus leading to full TCP connection. We can observe that the OpenFlow rule (3) has $n\_packets=0$.

	\begin{equation}
	\frac{Flow(ct\_state=+new, n\_packets=1400)}{Flow(ct\_state=+est, n\_packets=0)} \geq \delta
	\end{equation}
	\noindent The SDFW \textit{Local DFW-Event-Listener}, realizes that the threshold set for DDoS detection $\delta$ has been exceeded as shown above, and installs a new Flow rule with higher priority than the existing rule which allows TCP SYN packets, as shown below.
	
	\begin{lstlisting}
	ovs-ofctl add-flow br100 \
	"table=0, priority=51, nw_src=10.0.3.102,nw_dst=10.0.3.103,tcp actions=drop"
	\end{lstlisting}
	
	\noindent The malicious devices can also send a connection request to the switching software in the data plane. If the switch consists of flow rule entry corresponding to the traffic pattern, traffic is forwarded out of the specific switch port. If the entry is missing (table-miss packets) the request is sent to the controller. A class of DoS attacks - data to control plane saturation attacks as discussed by Gao \textit{et al}~\cite{gao2018security} can forge the OpenFlow fields with random values, that will lead to table-miss event in the switch. When a large volume of forged table-miss flows is sent to the controller as packet\_in entries. The controller can be saturated since these packet\_in messages will consume a large amount of switch-controller bandwidth and controller resources (CPU, memory). SDFW helps in the detection and mitigation of such attacks using state-based traffic analytics as discussed in the case study above.
	

	\section{Implementation and Evaluation}
	
	\subsection{Experimental Setup}
	We utilized an OpenStack based cloud network comprising of two Dell R620 servers and two Dell R710 servers all hosted in the data center. Each Dell server has about 128 GB of RAM and 16 core CPU. The SDN controller Opendaylight-Carbon was provided network management and orchestration in our framework. 
	
	\begin{table}[htb]
		\centering
		\caption{SDFW Components used in Implementation}
		\label{tab:5}
		\begin{tabular}{ p{2cm}|p{2cm}|p{2.5cm} }
			\hline  
			\textbf{Component} &     \textbf{LOC/Version} &     \textbf{Language / Framework}\\
			\hline\hline
			&& \\
			SDN Controller & OpenDaylight Carbon & Java, REST APIs \\
			\hline
			Local DFW-Event-Listener & 500 & python \\
			\hline
			Policy-Graph & 500 & python with Flask APIs\\
			\hline
			Flow Conflict Analyzer & 700 & python, networkx \\
			\hline
			Flow-Visualizer & 250 & python, d3, REST APIs \\
			\hline
			Data-Plane & 200 & Linux container LXC-3.0 \\
			\hline
			Frontend/UI & 400 & php-lavarel  \\
			\hline
		\end{tabular}                
	\end{table}
	
	\noindent In addition to these components - Table~\ref{tab:5}, we used the latest version of Open vSwitch (OVS 2.9.0), with conntrack module enabled to support the data plane connection tracking. 
	
	\subsection{SDFW Scalability Analysis}
	We conducted a scalability analysis to check the performance of SDFW when handling the TCP-SYN Flood attack. We conducted two separate experiments, one on a single switch topology, and one on a tree topology.  
	
	\begin{figure}[ht!]
		\centering
		\includegraphics[width=0.50\textwidth]{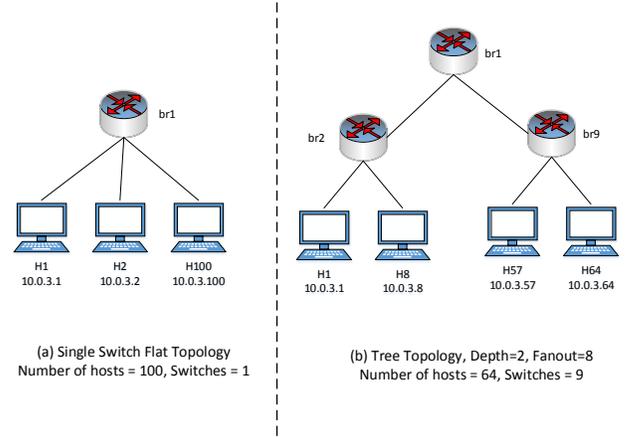}
		
		\caption{SDFW Scalability Analysis Experiment}
		\label{fig:10}
	\end{figure}
	\textbf{Flat Topology:} We utilized python script to run DDoS script from host H1 - 10.0.3.1 to perform TCP SYN-Flood on host H100 (10.0.3.100) - Figure~\ref{fig:10}(a). The experimental results show that once attack pattern is detected by SDFW, the attack-mitigation is enforced using OpenFlow rule on corresponding OpenFlow switch.
	We have currently utilized countermeasure to 'DROP' traffic flows for malicious traffic pattern, however other possible countermeasures include 'Rate-Limiting' the traffic flow, or redirecting to a honeypot for performing fine-grained packet analysis. 
	
	\begin{table}[htb]
		\centering
		\caption{SDFW Scalability DDoS Protection - Flat Topology}
		\label{tab:3}
		\begin{tabular}{ p{0.8cm}|p{1.3 cm}|p{1.3 cm}|p{1.3 cm}|p{1.3 cm} }
			\hline  
			\textbf{Hosts} & \textbf{BW-No-SDFW (Gb/s)} & \textbf{BW-SDFW (Gb/s)} & \textbf{Latency-No-SDFW (ms)}  & \textbf{Latency-SDFW (ms)}\\
			\hline 
			\hline
			100    & 42.96    & 37.6    & 8.76     & 9.66\\
			\hline
		\end{tabular}                
	\end{table}
	\noindent The experimental results - Table~\ref{tab:3} show that there is an 11\% drop in the network bandwidth - from  when using SDFW, and 9\% increase in the latency, i.e., from 8.76 ms to 9.66 ms when utilizing SDFW to inspect the connection state of network hosts. The drop in performance can be attributed to the fact that a single switch is receiving requests from about 100 hosts, and connection state of each host is analyzed using connection tracking module. 
	
	\textbf{Tree Topology:} The benefit of distributed firewall can be realized in a network having multiple switches, where each switch can locally track events from the hosts connected directly. We created a tree topology, with depth=2, fanout=8 in second experiment - Figure~\ref{fig:10}(b).  
	
	\begin{table}[htb]
		\centering
		\caption{SDFW Scalability DDoS Protection - Tree Topology}
		\label{tab:4}
		\begin{tabular}{ p{0.8cm}|p{1.3 cm}|p{1.3 cm}|p{1.3 cm}|p{1.3 cm} }
			\hline  
			\textbf{Hosts} & \textbf{BW-No-SDFW (Gb/s)} & \textbf{BW-SDFW (Gb/s)} & \textbf{Latency-No-SDFW (ms)}  & \textbf{Latency-SDFW (ms)}\\
			\hline 
			\hline
			64    & 35.9    & 35.3    & 8.87     & 9.2\\
			\hline
		\end{tabular}                
	\end{table}
	
	\noindent The experiment results - Table~\ref{tab:4}, show that in a network, with each switch checking attack-pattern for DDoS locally, then we have limited drop in performance. The bandwidth is reduced from 35.9 Gb/s to 35.4 Gb/s $\sim$ 1.6\% drop, which is acceptable for a moderate size network. Similarly, the network latency increases from 8.87 ms to 9.2 ms, when using SDFW for detecting the SYN-Flood attack, a $\sim$ 3.5\% increase. This gain in performance can be attributed to the fact, that benefit of distributed firewall implementation is obtained when multiple switches are involved. The experiments prove, that, in comparison to a centralized firewall model, the distributed stateful-firewall is able to scale well on a large network.
	
	\section{Related Work}
	\textbf{Distributed Stateful-SDN Security} is required to deal with attacks originating in SDN data-pane, as discussed by Bosshart \textit{et al}~\cite{bosshart2013forwarding}. The most relevant work to ours is what  Openstate~\cite{bianchi2014openstate} extended the OpenFlow switch to define a state-transition variable and extended finite-state-machine (XFSM) table, which is able to handle scenarios such as port knocking and TCP SYN-ACK message verification. The design is however based on centralized firewall architecture. The SDFW presented in this paper captures the recommendations defined in NIST 800-125b~\cite{chandramouli2016secure}  for protecting workloads within the data-center using next-generation distributed firewall (NDFW) model. Onix~\cite{koponen2010onix} uses distributed control plane design for SDN environment. We have used similar design principles for SDFW such as distributed virtual switch and network information base (NIB). P4~\cite{bosshart2014p4} is programming language which allows, protocol independent packet processing, and stateful packet inspection. We plan to extend the current work and develop a programming platform based on distributed firewall architecture.
	
	\section{Conclusion and Future Work}
	\noindent In this paper, we address the issue of security issues associated with lateral movement of attacker along the east-west plane in a data center, and packet flooding based data plane attacks. One limitation of this work is that we utilize SDFW to showcase defense against layer 4 security attacks. However, a next-generation firewall can also act as an application firewall and Deep-Packet-Inspection (DPI) module. As a part of future work, we plan to extend SDFW and address security attacks at the application layer.

	\ifCLASSOPTIONcompsoc
	\section*{Acknowledgments}
	\else
	\section*{Acknowledgment}
	\fi
	
	This research is based upon work supported by the NRL N00173-15-G017, NSF Grants 1642031, 1528099, and 1723440, and NSFC Grants 61628201 and 61571375.
	
	\ifCLASSOPTIONcaptionsoff
	\newpage
	\fi

	
	
	\bibliographystyle{IEEEtran}
	\bibliography{template}

\begin{thebibliography}{10}
\providecommand{\url}[1]{#1}
\csname url@samestyle\endcsname
\providecommand{\newblock}{\relax}
\providecommand{\bibinfo}[2]{#2}
\providecommand{\BIBentrySTDinterwordspacing}{\spaceskip=0pt\relax}
\providecommand{\BIBentryALTinterwordstretchfactor}{4}
\providecommand{\BIBentryALTinterwordspacing}{\spaceskip=\fontdimen2\font plus
\BIBentryALTinterwordstretchfactor\fontdimen3\font minus
  \fontdimen4\font\relax}
\providecommand{\BIBforeignlanguage}[2]{{%
\expandafter\ifx\csname l@#1\endcsname\relax
\typeout{** WARNING: IEEEtran.bst: No hyphenation pattern has been}%
\typeout{** loaded for the language `#1'. Using the pattern for}%
\typeout{** the default language instead.}%
\else
\language=\csname l@#1\endcsname
\fi
#2}}
\providecommand{\BIBdecl}{\relax}
\BIBdecl

\bibitem{kreutz2015software}
D.~Kreutz, F.~M. Ramos, P.~E. Verissimo, C.~E. Rothenberg, S.~Azodolmolky, and
  S.~Uhlig, ``Software-defined networking: A comprehensive survey,''
  \emph{Proceedings of the IEEE}, vol. 103, no.~1, pp. 14--76, 2015.

\bibitem{junipersdn}
J.~Networks, ``{Readiness, benefits, and barriers: An SDN progress report,},''
  \url{https://www.usebackpack.com/resources/7178/download?1451715494}, 2014,
  [Online; accessed 01-August-2018].

\bibitem{scott2013sdn}
S.~Scott-Hayward, G.~O'Callaghan, and S.~Sezer, ``Sdn security: A survey,'' in
  \emph{Future Networks and Services (SDN4FNS), 2013 IEEE SDN For}.\hskip 1em
  plus 0.5em minus 0.4em\relax IEEE, 2013, pp. 1--7.

\bibitem{mammela2016towards}
O.~M{\"a}mmel{\"a}, J.~Hiltunen, J.~Suomalainen, K.~Ahola, P.~Mannersalo, and
  J.~Vehkaper{\"a}, ``Towards micro-segmentation in 5g network security.''

\bibitem{hu2014flowguard}
H.~Hu, W.~Han, G.-J. Ahn, and Z.~Zhao, ``Flowguard: building robust firewalls
  for software-defined networks,'' in \emph{Proceedings of the third workshop
  on Hot topics in software defined networking}.\hskip 1em plus 0.5em minus
  0.4em\relax ACM, 2014, pp. 97--102.

\bibitem{porras2012security}
P.~Porras, S.~Shin, V.~Yegneswaran, M.~Fong, M.~Tyson, and G.~Gu, ``A security
  enforcement kernel for openflow networks,'' in \emph{Proceedings of the first
  workshop on Hot topics in software defined networks}.\hskip 1em plus 0.5em
  minus 0.4em\relax ACM, 2012, pp. 121--126.

\bibitem{suh2014building}
M.~Suh, S.~H. Park, B.~Lee, and S.~Yang, ``Building firewall over the
  software-defined network controller,'' in \emph{Advanced Communication
  Technology (ICACT), 2014 16th International Conference on}.\hskip 1em plus
  0.5em minus 0.4em\relax IEEE, 2014, pp. 744--748.

\bibitem{zerkane2016software}
S.~Zerkane, D.~Espes, P.~Le~Parc, and F.~Cuppens, ``Software defined networking
  reactive stateful firewall,'' in \emph{IFIP International Information
  Security and Privacy Conference}.\hskip 1em plus 0.5em minus 0.4em\relax
  Springer, 2016, pp. 119--132.

\bibitem{dixit2018challenges}
V.~H. Dixit, S.~Kyung, Z.~Zhao, A.~Doup{\'e}, Y.~Shoshitaishvili, and G.-J.
  Ahn, ``Challenges and preparedness of sdn-based firewalls,'' in
  \emph{Proceedings of the 2018 ACM International Workshop on Security in
  Software Defined Networks \& Network Function Virtualization}.\hskip 1em plus
  0.5em minus 0.4em\relax ACM, 2018, pp. 33--38.

\bibitem{gao2018security}
S.~Gao, Z.~Li, B.~Xiao, and G.~Wei, ``Security threats in the data plane of
  software-defined networks,'' \emph{IEEE Network}, 2018.

\bibitem{east-west}
\BIBentryALTinterwordspacing
Cisco, ``Trends in data center security,'' May 2014. [Online]. Available:
  \url{https://blogs.cisco.com/security/trends-in-data-center-security-part-1-traffic-trends}
\BIBentrySTDinterwordspacing

\bibitem{wack2002guidelines}
J.~Wack, K.~Cutler, and J.~Pole, ``Guidelines on firewalls and firewall
  policy,'' BOOZ-ALLEN AND HAMILTON INC MCLEAN VA, Tech. Rep., 2002.

\bibitem{shin2013avant}
S.~Shin, V.~Yegneswaran, P.~Porras, and G.~Gu, ``Avant-guard: Scalable and
  vigilant switch flow management in software-defined networks,'' in
  \emph{Proceedings of the 2013 ACM SIGSAC conference on Computer \&
  communications security}.\hskip 1em plus 0.5em minus 0.4em\relax ACM, 2013,
  pp. 413--424.

\bibitem{hunt2010zookeeper}
P.~Hunt, M.~Konar, F.~P. Junqueira, and B.~Reed, ``Zookeeper: Wait-free
  coordination for internet-scale systems.''

\bibitem{rosen2014linux}
R.~Rosen, ``Linux containers and the future cloud,'' \emph{Linux J}, vol. 240,
  no.~4, pp. 86--95, 2014.

\bibitem{bosshart2013forwarding}
P.~Bosshart, G.~Gibb, H.-S. Kim, G.~Varghese, N.~McKeown, M.~Izzard, F.~Mujica,
  and M.~Horowitz, ``Forwarding metamorphosis: Fast programmable match-action
  processing in hardware for sdn,'' in \emph{ACM SIGCOMM Computer Communication
  Review}, vol.~43, no.~4.\hskip 1em plus 0.5em minus 0.4em\relax ACM, 2013,
  pp. 99--110.

\bibitem{bianchi2014openstate}
G.~Bianchi, M.~Bonola, A.~Capone, and C.~Cascone, ``Openstate: programming
  platform-independent stateful openflow applications inside the switch,''
  \emph{ACM SIGCOMM Computer Communication Review}, vol.~44, no.~2, pp. 44--51,
  2014.

\bibitem{chandramouli2016secure}
R.~Chandramouli and R.~Chandramouli, ``Secure virtual network configuration for
  virtual machine (vm) protection,'' \emph{NIST Special Publication}, vol. 800,
  p. 125B, 2016.

\bibitem{koponen2010onix}
T.~Koponen, M.~Casado, N.~Gude, J.~Stribling, L.~Poutievski, M.~Zhu,
  R.~Ramanathan, Y.~Iwata, H.~Inoue, T.~Hama \emph{et~al.}, ``Onix: A
  distributed control platform for large-scale production networks.''

\bibitem{bosshart2014p4}
P.~Bosshart, D.~Daly, G.~Gibb, M.~Izzard, N.~McKeown, J.~Rexford,
  C.~Schlesinger, D.~Talayco, A.~Vahdat, G.~Varghese \emph{et~al.}, ``P4:
  Programming protocol-independent packet processors,'' \emph{ACM SIGCOMM
  Computer Communication Review}, vol.~44, no.~3, pp. 87--95, 2014.

\end{thebibliography}
	%
	
	
	
	%
	
	\begin{IEEEbiography}[{\includegraphics[width=1in,height=1.25in,clip,keepaspectratio]{ankur}}]{Ankur Chowdhary}
		is a Ph.D. Student in Computer Science at Arizona State University, Tempe, AZ, USA. He received B.Tech in Information Technology from GGSIPU in 2011 and MS in Computer Science from ASU in 2015. He has worked as Information Security Researcher for Blackberry Ltd., RSG and Application Developer for CSC Pvt. Ltd. His research interests include SDN, Web Security, Network Security and application of Machine Learning in the field of Security.
	\end{IEEEbiography}
	
	\begin{IEEEbiography}[{\includegraphics[width=1in,height=1.25in,clip,keepaspectratio]{dhuang8}}]{Dijiang Huang}
		received the B.S. degree from Beijing University of Posts and Telecommunications, Beijing, China, and the M.S. and Ph.D. degrees from the University of Missouri Kansas City, Kansas City, MO, USA, 1995, 2001, and 2004, respectively. He is an Associate Professor with the School of Computing Informatics and Decision System Engineering, Arizona State University, Tempe, AZ, USA. His research interests include computer networking, security, and privacy.	He is an Associate Editor of the Journal of Network and System Management (JNSM) and an Editor of the IEEE COMMUNICATIONS SURVEYS AND TUTORIALS. He has served as an organizer for many international conferences and workshops. His research was supported by the NSF, ONR, ARO, NATO, and Consortium of Embedded System (CES). He was the recipient of the ONR Young Investigator Program	(YIP) Award.
	\end{IEEEbiography}
	\begin{IEEEbiography}[{\includegraphics[width=1in,height=1.25in,clip,keepaspectratio]{adel1}}]{Adel Alshamrani}
		is a Ph.D. candidate in Computer Science at Arizona State University, Tempe, AZ, USA under the guidance of Dr. Dijiang Huang. He received his B.S. degree in computer science from Umm Al-Qura University, Saudi Arabia, and the M.S. degree in computer science from La Trobe University,
		Melbourne, Australia, in 2007 and 2010, respectively. He has 8 years of combined work experience in information security, network engineering, and teaching while working in the Faculty of Computing and Information Technology, King Abdul Aziz University. His research interests include information security, intrusion detection, and software defined networking.
	\end{IEEEbiography}
	
	
	
	
	

\end{document}